\documentclass{svjour3}                     % onecolumn (standard format)
\smartqed  % flush right qed marks, e.g. at end of proof
\usepackage{graphicx}
\usepackage{bm}
\usepackage[varg]{txfonts} % Times fonts
\usepackage[latin1]{inputenc}
\def\tri{{{}^3{\rm H}}}

\def\het{{{}^3{\rm He}}}
\def\heq{{{}^4{\rm He}}}

\def\be{\begin{equation}}
\def\ee{\end{equation}}
\def\bea{\begin{eqnarray*}}
\def\eea{\end{eqnarray*}}

\def\bi{\begin{itemize}}
\def\ei{\end{itemize}}

\journalname{Few-Body Systems (FB20)}
\begin{document}

\title{Effect of three nucleon forces in $p-\het$ scattering
%Four nucleon scattering states
\thanks{Presented at the 20th International IUPAP Conference on Few-Body Problems in Physics, 20 - 25 August, 2012, Fukuoka, Japan}
}


\author{
M. Viviani
\and %
L. Girlanda
\and %
A. Kievsky
\and %
L. E. Marcucci
}


\institute{%
M. Viviani \at INFN, Sezione di Pisa, Largo Pontecorvo, 3, 56127 Pisa
(Italy) \email{michele.viviani@pi.infn.it}
\and 
L. Girlanda \at Phys. Dept., Universit\`a del Salento, 73100 Lecce (Italy)
\\ and INFN Sezione di Lecce, 73100 Lecce (Italy) 
\and
A. Kievsky \at INFN, Sezione di Pisa, Largo Pontecorvo, 3, 56127 Pisa
(Italy)
\and
L.E. Marcucci \at 
Phys. Dept., University of Pisa,  Largo Pontecorvo, 3, 56127 Pisa
(Italy) and \\ INFN
 Sezione di Pisa, Largo Pontecorvo, 3, 56127 Pisa (Italy)
}

\date{Received: date / Accepted: date}

\maketitle

\begin{abstract}
The effect of the inclusion of different models of three nucleon (3N) forces in
$p-\het$ elastic scattering  at low energies
is studied. Two models have been considered: one derived from
effective field theory at next-to-next-to-leading 
order and one derived from a more phenomenological point of view -- the
so-called Illinois model. The four nucleon scattering observables are 
calculated using the Kohn variational principle and the
hyperspherical harmonic technique and the results are compared with
available  experimental data. We have found that with the
inclusion of both 3N force models the agreement with the experimental
data is improved, in particular for the proton vector analyzing 
power $A_{y}$.
\end{abstract}

\maketitle
%
% 
%----- Beginning of MAIN TEXT  --------------------------------------- 
% 
\section{Introduction}
\label{sec:intro}

Realistic nucleon-nucleon (NN) potentials reproduce the
experimental NN scattering data up to energies of 350 MeV
with a $\chi^2$ per datum close to 1. However, the use of these
potentials in the description of three-nucleon (3N) and four-nucleon (4N) bound
and scattering states gives a $\chi^2$ per datum much larger than
1 (see, e.g., Ref.~\cite{KVR01,Mea09}). To improve that situation, different
3N forces have been introduced~\cite{CS98}. 

Recently, the development of 3N forces has been brought forth following
mainly two lines. First, there are 3N force models derived from a
chiral effective field theory~\cite{Eea02}. At present, these models have been derived at
next-to-next-to-leading order (N2LO) of the so-called chiral
expansion. At this particular order, the 3N force contains two unknown
constants usually determined either by fitting the 3N and 4N
binding energies or, alternatively, the 3N binding energy and the
tritium $\beta$-decay Gamow-Teller matrix element~\cite{Mea12}. The 3N force depends
also on a cutoff function, 
which in general includes a cutoff parameter $\Lambda$. With a particular choice
of the cutoff function, a local version of the N2LO 3N interaction 
(hereafter referred as N-N2LO) has been derived~\cite{N07}. The parameter 
$\Lambda$ is chosen to be for physical reason of the order
of 500 MeV. The derivation of chiral 3N force at successive orders is
now in progress~\cite{BEKM11,KGE12}.  

Alternatively, within a more phenomenological approach, the so-called
Illinois model for the 3N force model has been
derived~\cite{PPWC01}. This model has been constructed to include
specific two- and three-pion exchange mechanisms between
the three nucleons. The model contains a few unknown parameters, which
have been determined by fitting the spectra of $A=4-12$ nuclei.  

It is clearly very important to test these new models in order to
understand how they can describe the nuclear dynamics. The
$A=3$ and $4$ scattering observables are between the best
testing grounds to understand their validity. It has been proven
that most of the $A=3$ scattering observables are quite insensitive to
the effect of the 3N force~\cite{KEMN12}. It is therefore of relevance to study their
effect in $A=4$ scattering observables.

In recent years, there has  been a rapid advance in solving the four nucleon
scattering problem with realistic Hamiltonians. Accurate calculations of four-body
scattering observables have been achieved in the framework of the
Alt-Grassberger-Sandhas (AGS) equations~\cite{DF07,DF07b}, solved in momentum
space, where the long-range Coulomb interaction is treated using the 
screening-renormalization me\-thod \cite{Alt78,DFS05}. Also
solutions of the Faddeev-Yakubovsky (FY) equations in configuration
space~\cite{Cie98,Lea05} and several calculations using the  
resonating group model \cite{PHH01,Sofia08}
were reported.  In this contribution, the four-body scattering problem
is solved using the Kohn variational method and expanding the internal
part of the wave function in terms of the hyperspherical harmonic (HH)
functions (for a review, see Ref. \cite{rep08}).  

Very recently, the efforts of the various groups have culminated 
in a benchmark paper~\cite{Vea11}, where it was shown that $p-\het$ and $n-\tri$ 
phase-shifts calculated using the AGS, FY, and HH techniques and using
several types of NN potentials are in very close agreement with each
other (at the level of less than 1\%).   

Since now the 4N scattering observables can be calculated with a good
accuracy, it is therefore timely to start to investigate the effect of
the 3N force in these systems. It is important to note that
the 4N studies performed so far have emphasized the presence of several
discrepancies between the theoretical predictions and experimental
data. Let us consider $p-\het$ elastic scattering, where exist several
accurate measurements of both the unpolarized cross
section~\cite{Fam54,Mcdon64,Fisher06}, the proton analyzing
power $A_y$~\cite{All93,Vea01,Fisher06}, and other polarization
observables~\cite{Dan10}. The calculations performed so
far with a variety of NN interactions have shown a glaring   
discrepancy between theory and experiment for
$A_y$~\cite{Fon99,Vea01,PHH01,Fisher06,DF07}. This discrepancy is very 
similar to the well known ``$A_y$  Puzzle''  in $N-d$ scattering. This is a
fairly old problem, already reported about 20 years ago~\cite{KH86,WGC88} in
the case of $n-d$ and later confirmed also in the $p-d$ case~\cite{Kie96}.
Also for other $p-\het$ observables (the $\het$ analyzing
power $A_{0y}$ and some spin correlation observables as $A_{yy}$,
$A_{xx}$, etc.) discrepancies have been observed. 

In this paper we report a preliminary study of the effect of including 3N forces
in $p-\het$ elastic scattering calculations in order to see if this
inclusion helps in reducing these discrepancies. Clearly, it is
important to specify which NN potential has been used together with a
particular model of 3N interaction. The 3N force derived from
the effective field theory at N2LO has been used together a NN
potential constructed within the same approach, in particular
the next-to-next-to-next-to-leading order (N3LO) interaction
by Entem and Machleidt~\cite{EM03}, with cutoff 
$\Lambda=500$ MeV (I-N3LO interaction model). 
The two free parameters of the N-N2LO 3N potential have been chosen
from the combination that reproduces the $A=3,4$ binding
energies~\cite{N07}. The Illinois 3N models has been used in
conjunction with the Argonne $v_{18}$ (AV18) NN potential~\cite{Wea95}. Between the
different Illinois models, we have considered the most recent one, the
so called Illinois-7 model (IL7)~\cite{il7}.

This paper is organized as follows. In Section~\ref{sec:res}, we
discuss the preliminary results obtained with four interaction models, two (I-N3LO
and AV18) including only NN interaction and two (I-N3LO/N-N2LO and
AV18/IL7) including also the 3N forces under consideration.
The conclusions will be given in Section~\ref{sec:conc}.

\begin{table}
\caption[Table]{\label{table:comp}
Phase-shifts and mixing angle parameters for $p-\het$ elastic
scattering at incident proton energy $E_p=5.54$ MeV calculated using the
I-N3LO potential. The values reported in the columns labeled HH have been
obtained using the HH expansion and the Kohn variational principle,
those reported in the columns labeled AGS by solving the AGS
equations~\protect\cite{DF07b}, and those reported in the columns
labeled FY by solving the FY
equations~\protect\cite{Lea05}.
}
\begin{center}
\begin{tabular}{l@{$\quad$} c@{$\quad$}c@{$\quad$}c@{$\quad$}|
                l@{$\quad$}c@{$\quad$}c@{$\quad$}c}
\hline
Phase-shift & HH  &  AGS  & FY   & Phase-shift & HH  & AGS  & FY   \\
\hline
${}^1S_0$  & $-68.5$ & $-68.3$ & $-69.0$   &
${}^3P_0$  & $25.1$  & $25.4$  & $25.8$   \\
\hline
${}^3S_1$  & $-60.1$ & $-60.0$ & $-60.0$   & 
${}^1P_1$  & $23.0$  & $23.0$  & $23.2$ \\
${}^3D_1$  & $-1.51$ & $-1.45$ & $-1.40$ & 
${}^3P_1$  & $44.3$  & $44.5$  & $44.1$  \\
$\epsilon(1^+)$ & $-1.07$ & $-1.08$ & $-1.18$ & 
$\epsilon(1^-)$ & $9.36$  & $9.28$  & $9.28$ \\
\hline
\end{tabular}
\end{center}
\end{table}

\section{Study of the 3N force}
\label{sec:res}
In this Section, we present the results of the inclusion of the
considered 3N force models in elastic $p-\het$ observables. 
This work is still preliminary, final results with complete tests of
numerical stability will be reported in a forthcoming
paper~\cite{newHH}. We present in particular the results obtained
using I-N3LO, AV18, I-N3LO/N-N2LO, and AV18/IL7 models. 

The HH method and the Kohn variational principle used to perform the
calculations are described in Ref.~\cite{rep08}. 
The accuracy reached by this technique is rather good, as evidenced
in Ref.~\cite{Vea11}, where $p-\het$ and $n-\tri$ phase shifts 
calculated with the present method were found in good agreement with
the results obtained by other groups using the AGS~\cite{DF07b} and
FY~\cite{Lea05} techniques. 
As an example, a comparison of a selected set of phase-shifts
and mixing angle parameters for $p-\het$ elastic scattering at
$E_p=5.54$ MeV calculated by means of the HH, AGS, and
FY techniques is reported in Table~\ref{table:comp}. As
can be seen, there is a good overall agreement between the results of
the two calculations. 

\begin{center}
\begin{figure}[tb]
 \includegraphics[clip,width=\columnwidth]{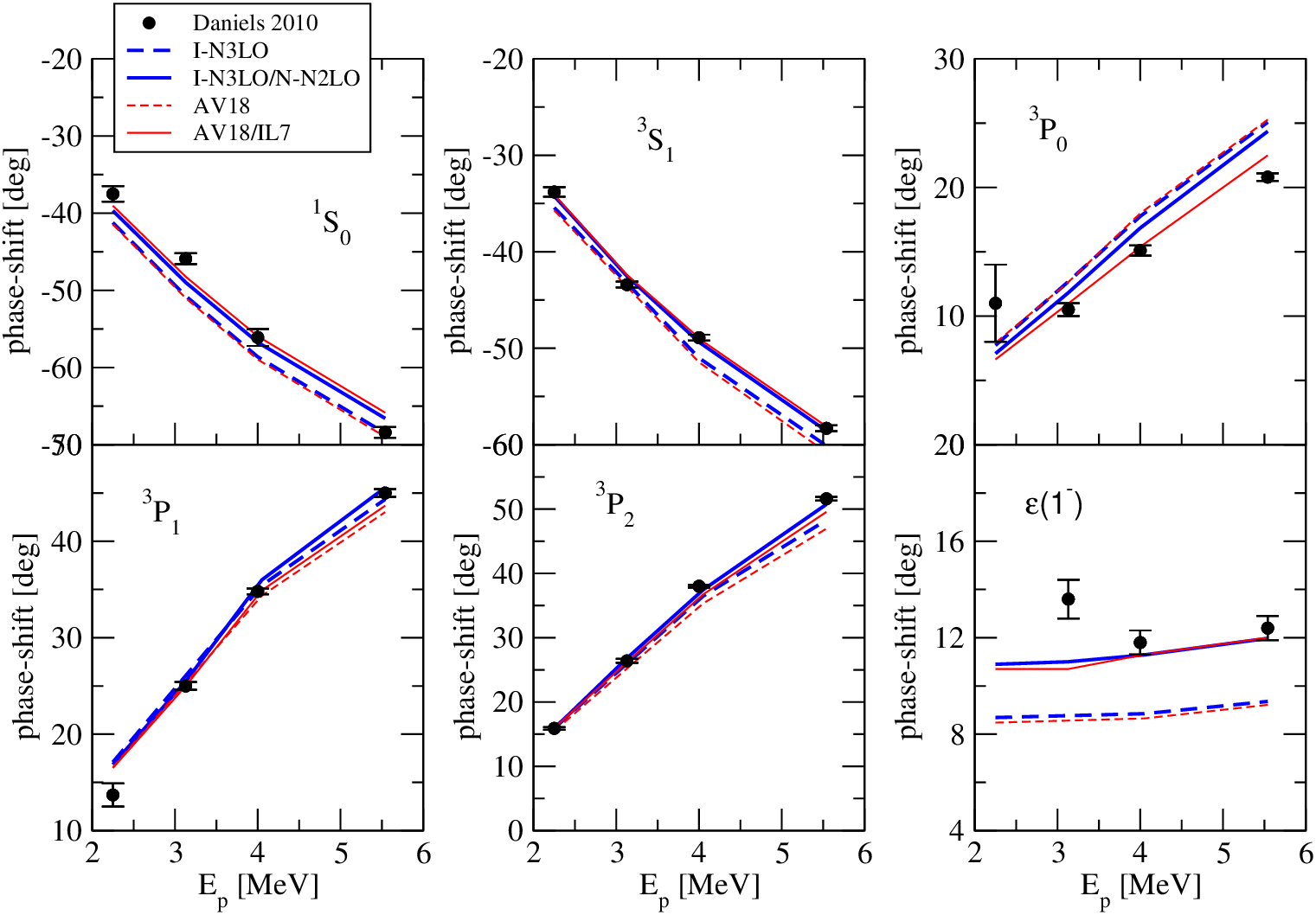}
\caption{$p-\het$ phase shifts calculated with  I-N3LO (blue dashed
  line), I-N3LO/N-N2LO (blue solid line),  AV18 (thin red dashed
  line), and  AV18/IL7 (thin red solid line) interaction models. The results 
  of the PSA performed at TUNL have been also reported~\protect\cite{Dan10}.}
\label{fig:psa}
\end{figure}
\end{center}

\begin{figure*}[!tb]
 \includegraphics[clip,width=\columnwidth]{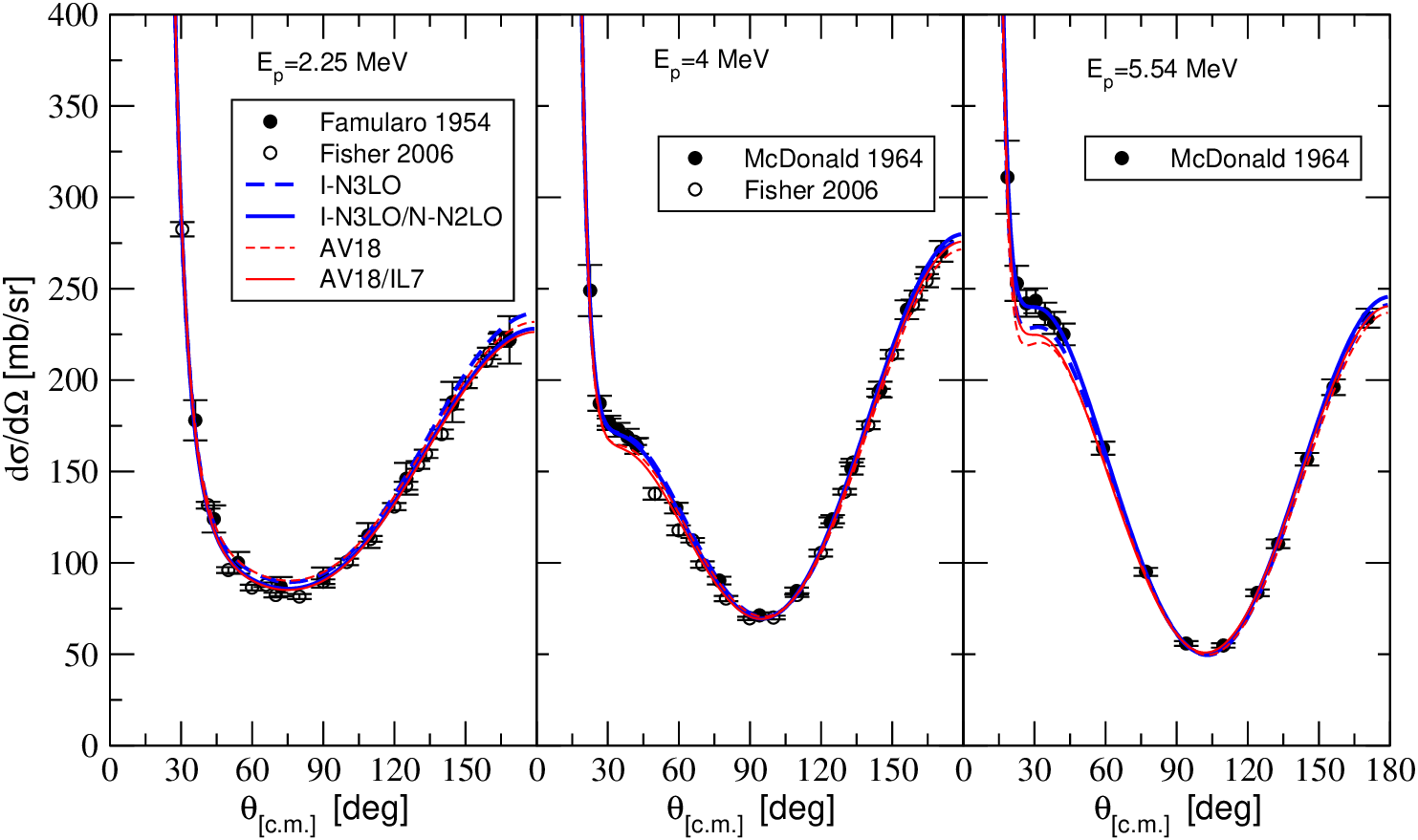}
 \caption{$p-\het$ differential cross sections
 for three  different incident proton 
  energies (notation as in Fig.~\protect\ref{fig:psa}). The
  experimental data are from 
  Refs.~\protect\cite{Fam54,Mcdon64,Fisher06}. }  
\label{fig:dcs}
\end{figure*}

In the energy range considered here ($E_p\le 6$ MeV), the various
$p-\het$ observables are dominated by S-wave and P-wave phase shifts
(D-wave phase shifts give only a marginal contribution, and more
peripheral phase shifts are negligible). A comparison of a
selected set of calculated phase-shifts and mixing parameters with those
obtained by the recent PSA~\cite{Dan10} is reported in
Fig.~\ref{fig:psa}. For the ${}^1S_0$ and ${}^3S_1$ phase shifts, the results 
obtained including NN interactions slightly overpredict
(in absolute value) the PSA values. Including the 3N force,
the calculated phase shifts agree very well with the PSA values (for
both AV18/IL7 and I-N3LO/N-N2LO models). In fact, the $p-\het$
interaction in S-wave is repulsive, being dominated by the Pauli repulsion,
and the corresponding phase shifts are generally well reproduced by an
interaction model giving the correct value of the $\het$ binding energy (and
radius).

Let us consider now P-wave phase shifts. For the ${}^3P_0$ phase shift we observe
that the models including NN interaction only overpredict the PSA
values. With the inclusion of the 3N forces, the results come close to
the PSA values. Note that the  ${}^3P_0$ phase shift calculated with
the I-N3LO/N-N2LO model still overpredict the PSA values, while the
values obtained with AV18/IL7 are in better agreement with data.  The 
${}^3P_2$ and ${}^3P_1$ phase shifts are underpredicted by the models
including NN interaction only, and again including the 3N force the
agreement with the PSA values is improved. For these two phase shifts,
the values calculated with the I-N3LO/N-N2LO models are in slightly better
agreement with the PSA values. Finally, for the $1^-$ mixing parameter
we observe that the theoretical results obtained with the I-N3LO and AV18 models
disagree with the PSA values. On the contrary, the results obtained
with AV18/IL7 and I-N3LO/N-N2LO are very close to the data. In
conclusion, including a 3N force, we observe a general improvement 
of the description of the S-- and P--wave phase shifts and mixing
parameters.

Let us now compare the theoretical results directly with a
selected set of available experimental data (the complete
comparison will be presented in the forthcoming paper~\cite{newHH}). The calculated
$p-\het$ differential cross sections a energies $E_p=2.25$, $4$, and
$5.54$ MeV are reported in Fig.~\ref{fig:dcs} and compared with the
experimental data of Refs.~\cite{Fam54,Mcdon64,Fisher06}. As can be
seen, there is a good agreement between the theoretical calculations 
and experimental data. In fact, this observable is sensitive to small
changes of the phase shifts only in the ``interference'' region around
$\theta=30$ deg. We note that only the I-N3LO/N-N2LO reproduces
the differential cross section there, as is more evident at
$E_p=5.54$ MeV.

\begin{figure*}[htb]
 \includegraphics[clip,width=\columnwidth]{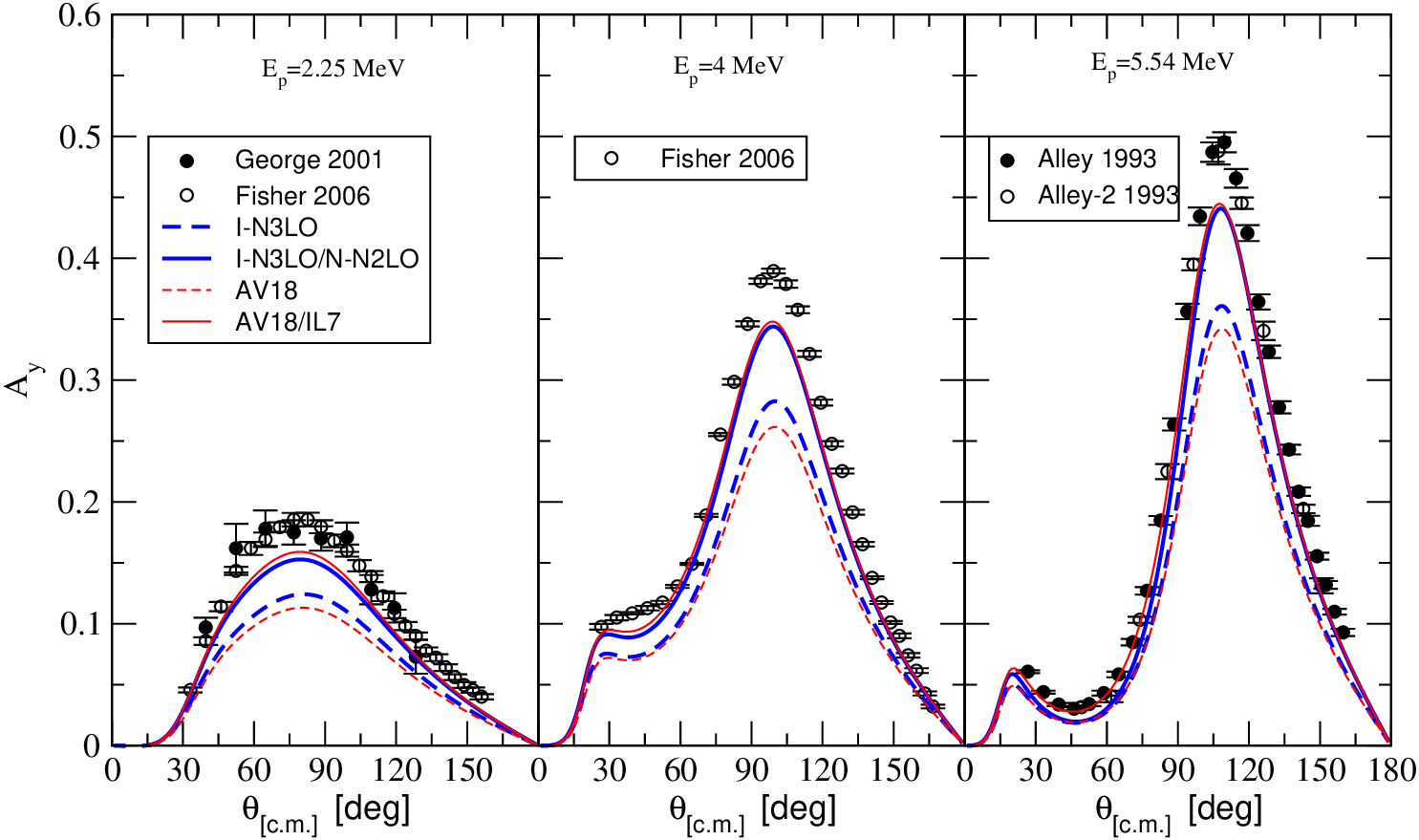}
 \caption{Same as in Fig.~\protect\ref{fig:dcs} but for the
  spin-correlation $A_y$ observable. The experimental data are from
  Refs.~\protect\cite{All93,Vea01,Fisher06}. }  
\label{fig:ay}
\end{figure*}

More interesting is the situation for the proton vector analyzing power
$A_y$, shown in Fig.~\ref{fig:ay}. Here, we observe a larger sensitivity to the employed
interaction model. The calculations performed using I-N3LO and AV18
largely underpredict the experimental points, a fact already observed
before~\cite{Vea01,Fisher06,Vea11}. A sizable improvement is found by
adopting both I-N3LO/N-N2LO and AV18/IL7 models, as it was expected from 
the discussion regarding the comparison with the PSA phase shifts. 
The analyzing powers calculated including the two
3N force models are very close with each other.

\begin{figure}[t]
 \includegraphics[clip,width=\columnwidth]{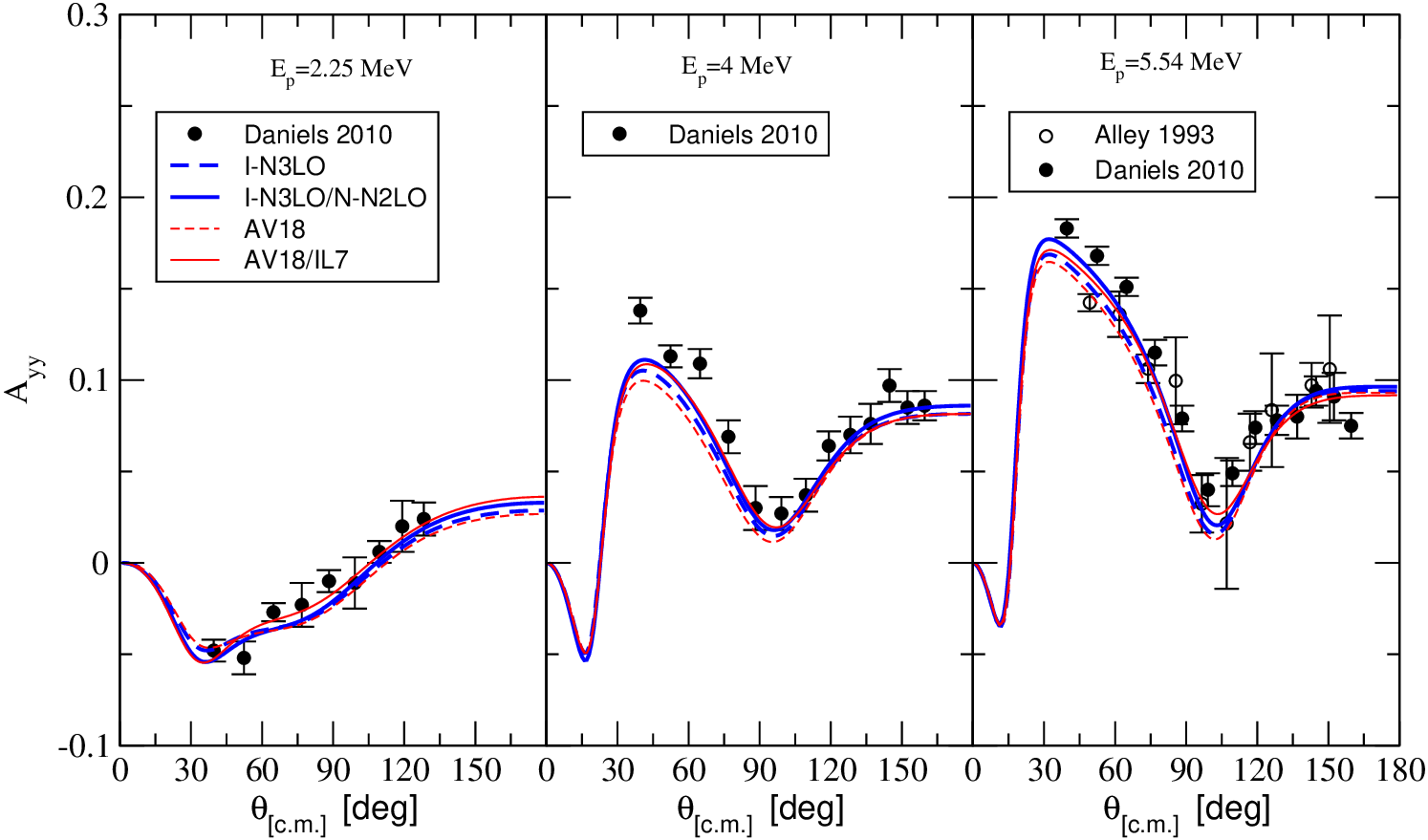}
 \caption{Same as in Fig.~\protect\ref{fig:dcs} but for the
  spin-correlation $A_{yy}$ observable. The experimental data
  are from Refs.~\protect\cite{All93,Dan10}. } 
\label{fig:ayy}
\end{figure}

Finally, in Fig.~\ref{fig:ayy}, we show a further polarization
observable, the $\het$ spin correlation coefficient $A_{yy}$. This
observable (and other spin correlation coefficients) is found to be
not very sensitive to the interaction model. We observe however that the
I-N3LO/N-N2LO results are in slightly better agreement with data.

\section{Conclusions}
\label{sec:conc}

In this paper preliminary results for $p-\het$ elastic scattering
including the effect different of 3N force models have been reported. We have
considered two 3N interaction models. The IL7 model is based on
a phenomenological approach, being
constructed from a select set of two- and three-pion
exchange processes between the three nucleons. The N-N2LO model is derived from
a chiral effective theory up to N2LO in the chiral expansion.
Both models used in conjunction with a ``consistent'' NN potential
(AV18 with IL7 and I-N3LO with N-N2LO) reproduce the $A=3,4$
binding energies. 

The results obtained for the considered scattering observables 
have been compared with the available experimental data and a very
recent PSA performed at TUNL~\cite{Dan10}. We have found that the 
phase shifts obtained with both the I-N3LO/N-N2LO and AV18/IL7 models are very close
with those derived from the PSA. The direct comparison of the
theoretical results with the experimental data has shown that there
are still some discrepancies, but the $A_y$ problem is noticeably
reduced. In fact, we observe that now the discrepancy is reduced to be
of the order of 10\% at the peak, much less than before. 
We have also found that the results obtained with the I-N3LO/N-N2LO
and AV18/IL7 are always very close with each other (see Fig.~\ref{fig:ay}). Since the
frameworks used to derive these 3N force models are rather
different, this outcome is somewhat surprising. 

Finally, we would like to remark that the previously observed large underprediction of
the $p-\het$ $A_y$ observable was considered to be due to some deficiencies
of the interaction in P-waves~\cite{Fon99,Vea01}, as, for example,
due to the appearance of a unconventional ``spin-orbit'' interaction
in $A>2$ systems~\cite{K99}. In fact, some of the parameters of
the IL7 model have been fitted to reproduce the P-shell
nuclei spectra~\cite{il7}. Therefore, the AV18/IL7 model is 
constructed to take into account (at least effectively) of this
unconventional ``spin-orbit'' interaction and this can explain
the improvement in the description of the $p-\het$ $A_y$
observable. Regarding the N-N2LO 3N force model, its two parameters
have been fitted to the $A=3$ and $A=4$ binding energies. Therefore,
its capability to improve the description of the $p-\het$ $A_y$ observable is not
imposed but it is somewhat built-in. Investigations to understand these
issues are in progress. It is interesting to note that
in the $N-d$ case, the use of the I-N3LO/N-N2LO model
does not give a significant improvement in the solution of the
``$A_y$ puzzle''~\cite{Mea09}. A detailed analysis of effect of the
AV18/IL7 interaction in $A=3,4$ systems is currently
underway~\cite{newHH}. Finally, it will be certainly very 
interesting to test the effect of the inclusion of the next order 3N
forces derived from the effective field theory.

\end{document}